\documentclass[prb,a4paper,twocolumn,floatfix,superscriptaddress,amsmath,amsfonts,amssymb,citeautoscript,preprintnumbers,longbibliography]{revtex4-2}
\pdfoutput=1 \widowpenalty10000 \clubpenalty10000
\usepackage[T1]{fontenc}\usepackage[latin1]{inputenc}
\usepackage{dcolumn,graphicx,color,booktabs,microtype,afterpage,upgreek,fixmath,relsize} 
\usepackage[slantedGreek]{mathptmx}
\usepackage{textcomp,pzccal}

\newcount\hh \newcount\mm
\hh=\time \divide\hh by 60
\mm=\hh \multiply\mm by 60 \mm=-\mm
\advance\mm by \time
\def\now{\number\hh:\ifnum\mm<10{}0\fi\number\mm}

\usepackage{hyperref}
\hypersetup{colorlinks, plainpages=false, linkcolor=blue, urlcolor=blue, citecolor=blue, pdfpagemode=UseNone, pdfstartview=FitBH}

\newcommand{\pmmp}{^\text{\raisebox{-0.5pt}{\textpm}}{\scriptstyle\!\!\!\text{\raisebox{0.5pt}{$\diagup$}}\!\!\!}_\text{\rotatebox[origin=c]{180}{\raisebox{-0.5pt}{\textpm}}}\,}
\newcommand{\txtmp}{\rotatebox[origin=c]{180}{\textpm}}
\newcommand{\hamcal}{\kern3pt\hat{\kern-3pt\mathcal{H}}\kern-1pt}

\begin{document}

\makeatletter\renewcommand{\ps@plain}{%
\def\@evenhead{\hfill\itshape\rightmark}%
\def\@oddhead{\itshape\leftmark\hfill}%
\renewcommand{\@evenfoot}{\hfill\small{--~\thepage~--}\hfill}%
\renewcommand{\@oddfoot}{\hfill\small{--~\thepage~--}\hfill}%
}\makeatother\pagestyle{plain}

\newcommand{\CeThreePlus}{Ce$^\text{3+}$}
\newcommand{\CePdSi}{Ce$_\text{3}$Pd$_\text{20}$Si$_\text{6}$}


\title{Cascade of magnetic-field-driven quantum phase transitions in Ce$_3$Pd$_{20}$Si$_6$}

\author{F.~Mazza}
\affiliation{Institute of Solid State Physics, Vienna University of Technology, Wiedner Hauptstr. 8--10, 1040 Vienna, Austria}
\affiliation{Institut Laue-Langevin, 71 avenue des Martyrs CS 20156, 38042 Grenoble Cedex 9, France}

\author{P.~Y.~Portnichenko}
\affiliation{Institut f{\"u}r Festk{\"o}rper- und Materialphysik, Technische Universit{\"a}t Dresden, 01069 Dresden, Germany}

\author{S.~Avdoshenko}
\affiliation{\mbox{Leibniz-Institut f\"ur Festk\"orper- und Werkstoffforschung (IFW) Dresden, Helmholtzstra{\ss}e 20, 01069 Dresden, Germany}}

\author{P.~Steffens}
\affiliation{Institut Laue-Langevin, 71 avenue des Martyrs CS 20156, 38042 Grenoble Cedex 9, France}

\author{M.~Boehm}
\affiliation{Institut Laue-Langevin, 71 avenue des Martyrs CS 20156, 38042 Grenoble Cedex 9, France}

\author{\mbox{Eun Sang Choi}}
\affiliation{National High Magnetic Field Laboratory, Florida State University, Tallahassee, Florida 32310-3706, USA}

\author{M.~Nikolo}
\affiliation{Department of Physics, Saint Louis University, St.~Louis, Missouri 63103, USA}

\author{X.~Yan}
\affiliation{Institute of Solid State Physics, Vienna University of Technology, Wiedner Hauptstr. 8--10, 1040 Vienna, Austria}

\author{A.~Prokofiev}
\affiliation{Institute of Solid State Physics, Vienna University of Technology, Wiedner Hauptstr. 8--10, 1040 Vienna, Austria}

\author{S.~Paschen}
\affiliation{Institute of Solid State Physics, Vienna University of Technology, Wiedner Hauptstr. 8--10, 1040 Vienna, Austria}

\author{D.~S.~Inosov}\email[Corresponding author: \vspace{5pt}]{dmytro.inosov@tu-dresden.de}
\affiliation{Institut f{\"u}r Festk{\"o}rper- und Materialphysik, Technische Universit{\"a}t Dresden, 01069 Dresden, Germany}

\begin{abstract}\parfillskip=0pt\relax
\noindent
Magnetically hidden order is a hypernym for electronic ordering phenomena that are visible to macroscopic thermodynamic probes but whose microscopic symmetry cannot be revealed with conventional neutron or x-ray diffraction. In a handful of \textit{f}-electron systems, the ordering of odd-rank multipoles leads to order parameters with a vanishing neutron cross-section. Among them, Ce$_3$Pd$_{20}$Si$_6$ is known for its unique phase diagram exhibiting two distinct multipolar-ordered ground states (phases II and II$^\prime$), separated by a field-driven quantum phase transition associated with a putative change in the ordered quadrupolar moment from $O^0_2$ to~$O_{xy}$. Using torque magnetometry at subkelvin temperatures, here we find another phase transition at higher fields above 12~T, which appears only for low-symmetry magnetic field directions \mbox{$\mathbf{B}\parallel\langle11L\rangle$ with $1 < L \leq 2$}. While the order parameter of this new phase~II$^{\prime\prime}$ remains unknown, the discovery renders Ce$_3$Pd$_{20}$Si$_6$ a unique material with two field-driven phase transitions between distinct multipolar phases. They are both clearly manifested in the magnetic-field dependence of the field-induced $(111)$ Bragg intensities measured with neutron scattering for $\mathbf{B}\parallel[11\overline{2}]$. We also find from inelastic neutron scattering that the number of nondegenerate collective excitations induced by the magnetic field correlates with the number of phases in the magnetic phase diagram for the same field direction. Furthermore, the magnetic excitation spectrum suggests that the new phase~II$^{\prime\prime}$ may have a different propagation vector, revealed by the minimum in the dispersion that may represent the Goldstone mode of this hidden-order phase.
\end{abstract}


\maketitle

\section{Introduction}\vspace{-5pt}

The appearance of so-called hidden order phases in the low-temperature magnetic phase diagrams of \textit{f}-electron systems has been confronted with intensified interest in recent years~\cite{Buyers96, PaixaoDetlefs02, SantiniCarretta09, CameronFriemel16, ShenLiu19, KuramotoKusunose09}. Contrary to the conventional magnetic order composed of atomic spins, multipolar order parameters~\cite{KusunoseKuramoto01, Kusunose08, KuramotoKusunose09, YatsushiroKusunose21} involve both spin and orbital degrees of freedom, linked via strong spin-orbit coupling, and are highly sensitive to an applied magnetic field~\cite{ShiinaShiba97, ShiinaSakai98, PortnichenkoAkbari20, ThalmeierAkbari21, InosovAvdoshenko21}. Consequently, field-induced quantum phase transitions and quantum critical behavior emerge~\cite{StrydomPikul06, PaschenMueller07, CustersLorenzer12, Schofield10}, showing remarkable anisotropies with respect to the magnetic field direction~\cite{GotoWatanabe09, MitamuraTayama10, OnoNakano13, OkazakiShibauchi11, InosovAvdoshenko21}. In addition, hybridization of the \textit{f} electrons with the conduction band may lead to either heavy-fermion metal or Kondo-insulator states~\cite{CustersLorenzer12, MartelliCai19}.

Previous studies on a variety of 4\textit{f}- and 5\textit{f}-electron systems already demonstrated the presence of ordered phases with multipolar order parameters, for example in CeB$_{6}$~\cite{HanzawaKasuya84, CameronFriemel16}, NpO$_2$~\cite{PaixaoDetlefs02, PourovskiiKhmelevskyi21}, URu$_{2}$Si$_{2}$~\cite{ChandraColeman02, HauleKotliar09, MydoshOppeneer11, IkedaSuzuki12, ShibauchiIkeda14, ThalmeierTakimoto14} or filled skutterudites~\cite{TayamaSakakibara03, KuramotoKusunose09}. Identifying the symmetry of hidden order is not an easy task since common neutron and x-ray diffraction in zero field produce no signal. Nevertheless, some alternative techniques may be used to reveal the primary antiferroquadrupolar (AFQ) and field-induced dipolar- and octupolar states, such as resonant \cite{NakaoMagishi01, NagaoIgarashi01, NagaoIgarashi06, MatsumuraYonemura09} and nonresonant~\cite{TanakaStaub04, TanakaSera06} x-ray scattering, as well as neutron diffraction in an external field~\cite{ErkelensRegnault87}.

An alternative strategy is to look at the magnetic excitations spectrum, with the objective of unraveling the propagation vector of the hidden-order phase indirectly from a minimum in the dispersion relations~\cite{ShenLiu19, PortnichenkoNikitin19}. In the presence of a suitable theoretical model describing the magnetic excitations, the order parameter and multipolar interactions can in principle be extracted from a comparison of inelastic neutron scattering (INS) data with the model, yet the main challenge is the large number of free parameters describing the RKKY-type couplings between different multipoles across several nearest neighbors \cite{YamadaHanzawa19, HanzawaYamada19}. Here one may benefit from considering the anisotropy of field-induced excitations in field space, as demonstrated recently for CeB$_6$~\cite{PortnichenkoAkbari20}. The dynamical structure factor $\textit{S}(\textbf{\textit{Q}},\omega)$, measured by single-crystal INS in high magnetic fields applied along different crystal axes, was shown to be fairly consistent with theoretical results obtained with Holstein-Primakoff (HP)~\cite{ShiinaShiba03} and random-phase approximation (RPA) methods~\cite{ThalmeierShiina98, ThalmeierShiina03, ThalmeierShiina04, ThalmeierAkbari21}.

The cage compound Ce$_3$Pd$_{20}$Si$_6$ with a cubic $Fm\overline{3}m$ structure is famous as a unique system that shows two distinct multipolar-ordered phases separated by a field-driven quantum critical point, where thermodynamic and transport measurements revealed remarkable non-Fermi-liquid behavior~\cite{MartelliCai19}. In zero field, this compound exhibits an antiferromagnetic (AFM) phase~III below \textit{T}$_{\rm N}$ = 0.31~K~\cite{GotoWatanabe09, MitamuraTayama10, OnoNakano13} with a propagation vector $q_\text{III}=(0\,0\,\frac{4}{5})$, followed in temperature by an antiferroquadrupolar (AFQ) phase~II below \textit{T}$_{\rm Q}$ = 0.5~K. The propagation vector of the AFQ phase was previously identified by the appearance of field-induced magnetic satellites at a slightly incommensurate position close to the $(111)$ structural reflection, where in zero field only diffuse dynamical fluctuations are seen~\cite{PortnichenkoCameron15, PortnichenkoPaschen16}. Due to the field-induced dipolar moments, the magnetic peaks became visible to neutrons, confirming the ordering of $O^0_2$-type quadrupoles~\cite{PortnichenkoPaschen16} in agreement with the proposed theory~\cite{ShiinaShiba97, ShiinaSakai98, ShiinaShiba03}.

When magnetic field is applied to the sample, the stability region of phase~II strongly depends on the field direction. According to magnetization, specific-heat, and ultrasound measurements~\cite{GotoWatanabe09, MitamuraTayama10, OnoNakano13}, it persists to approximately 18~T for $\mathbf{B}\parallel [111]$ and up to 10~T for $\mathbf{B}\parallel [110]$. However, for $\mathbf{B}\parallel [001]$, phase~II is suppressed already at 2~T, where a phase transition to the second, field-induced AFQ phase~II$^\prime$ occurs. According to the group-theoretical analysis~\cite{ShiinaShiba97, ShiinaSakai98}, the $\Gamma_8$-quartet ground state of the Ce$^{3+}$ ion in a cubic crystal field supports different types of quadrupolar moments: $\Gamma_3^+$-type quadrupoles $O_2^0$ and $O_2^2$, and $\Gamma_5^+$-type quadrupoles $O_{xy}$, $O_{yz}$, and $O_{zx}$. An external magnetic field applied along some general direction given by the unit vector $(\alpha\beta\gamma)$ selects a coherent superposition of the initially degenerate quadrupoles within each manifold as the primary order parameter, e.g. $\alpha O_{yz}+\beta O_{zx}+\gamma O_{xy}$ for $\Gamma_5^+$, which depends continuously on the field direction~\cite{ThalmeierAkbari21}.

The vanishing field-induced Bragg intensity, observed across the transition from phase II to phase II$^\prime$ with neutron scattering~\cite{PortnichenkoPaschen16}, is consistent with the proposed change in the type of quadrupolar ordering from $O^{0}_2$ ($\Gamma_3^+$-type) in phase~II to $O_{xy}$ ($\Gamma_5^+$-type) in phase~II$^\prime$~\cite{CustersLorenzer12}, because due to symmetry considerations, the $O_{xy}$ moments support no field-induced dipoles for $\mathbf{B}\parallel [001]$~\cite{ShiinaShiba97}. A similar field-driven phase transition between AFQ phases with different ordered moments was first proposed theoretically for CeB$_6$~\cite{ShiinaShiba97}, yet it was never observed experimentally in this compound. Most likely, the weaker magnetic interactions in Ce$_3$Pd$_{20}$Si$_6$ that are responsible for a sevenfold reduction in the ordering temperatures compared to CeB$_6$ also drive the critical fields to values that can be easily reached in experiments, making it an ideal system for studying transitions between different multipolar phases. Two such transitions were later found in the cubic compound PrOs$_4$Sb$_{12}$ for $\mathbf{B}\parallel[110]$~\cite{TayamaSakakibara03}. More recently, a somewhat similar field-driven ``metamultipolar'' phase transition has also been revealed in the unconventional superconductor CeRh$_2$As$_2$ for magnetic fields applied in the tetragonal \textit{ab} plane~\cite{HafnerKhanenko22}.

The large field-direction anisotropy suggested by the magnetic phase diagrams of Ce$_3$Pd$_{20}$Si$_6$, acquired so far only for high-symmetry directions of the magnetic field, calls for a more systematic study of their field-angle dependence. In particular, measurements of phase~II$^\prime$ in fields rotated away from the $[001]$ axis are essential both for understanding the stability region of this phase and for confirming its underlying order parameter. Indeed, the symmetry constraints that lead to vanishing field-induced dipolar moments on the $O_{xy}$ quadrupoles apply only for $\mathbf{B}\parallel [001]$ but are relieved for any other arbitrary field axis~\cite{ShiinaShiba97}. It is therefore expected that field-induced magnetic Bragg peaks in phase~II$^\prime$ can be revealed by rotating the field axis away from $[001]$, as long as the phase itself is not suppressed by this rotation.

These considerations motivate our present study. In Sec.~\ref{Sec:Magnetometry} we present the low-temperature magnetic phase diagram of Ce$_3$Pd$_{20}$Si$_6$ in magnetic field space using torque magnetometry and specific heat. These measurements exhibit a weak anomaly, only visible for a limited range of field directions, that may evidence an additional high-field phase~II$^{\prime\prime}$ beyond phase~II$^\prime$. Equipped with the field-angle phase diagrams resulting from thermodynamic measurements, we then proceed in Sec.~\ref{Sec:Neutrons} to the results of our neutron scattering measurements for magnetic field $\mathbf{B}\parallel[11\overline{2}]$, where we directly reveal both multipolar phase transitions and the order parameter of phase~II$^\prime$. We also present the INS spectrum of collective multipolar excitations and analyze their dispersion and magnetic-field dependence, showing that additional field-induced collective modes appear at every phase transition separating different multipolar phases. Further, dispersion minima of the lowest-energy excitation hint at the candidate propagation vectors for the so far unidentified phase~II$^{\prime\prime}$.

\vspace{-6pt}\section{Single-ion properties of \protect\CeThreePlus\ in \protect\CePdSi}\vspace{-1pt}\label{Sec:Theory}\vspace{-5pt}

We start with introducing the crystal structure and magnetic properties of Ce ions in Ce$_3$Pd$_{20}$Si$_6$ at the single-ion level. This compound crystallizes in the $Fm\overline{3}m$ cubic space group with two crystallographically inequivalent Ce sites at the 4a (Ce1) and 8c (Ce2) Wyckoff positions~\cite{GribanovSeropegin94}. The Ce1 site is surrounded by 12 Pd atoms and 6 Si atoms, resulting in octahedral ($O_h$) point symmetry, whereas the Ce2 site is tetrahedrally coordinated with 16 Pd atoms ($T_d$ point symmetry)~\cite{JakovacHorvatic20}. Based on recent experimental evidence from diffuse neutron scattering, it was suggested that the coherent magnetic neutron-scattering intensity is dominated by the contribution from the simple-cubic Ce2 sublattice, whereas the face-centered-cubic Ce1 sublattice is magnetically silent~\cite{PortnichenkoCameron15}. We, therefore, used the DFT/PBE/PAW level of theory as implemented in the \texttt{VASP5} code (for details, see Appendix) to estimate the local charge rearrangements around the Ce2 site. We then proceed with \textit{ab initio} calculations of the crystal electric field (CEF) levels and their $|J,m_J\rangle$ compositions for the lowest-energy multiplet of Ce$^{3+}$ ($J=5/2$) on this site by employing the multi-configurational complete active space self-consistent field (CASSCF) method as implemented in the \texttt{OpenMolcas} code~\cite{RoosLindh08, AquilanteAutschbach20}.

\begin{figure}[b!]
\includegraphics[width=\columnwidth]{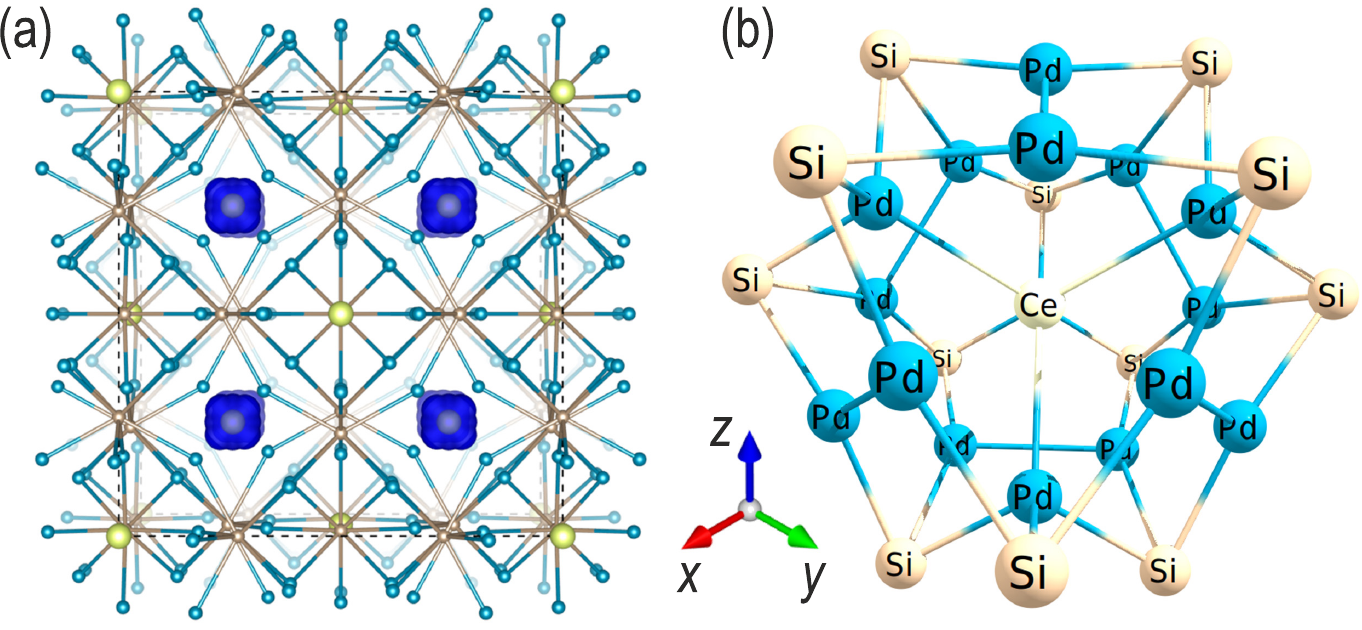}\vspace{-3pt}
\caption{(a)~The crystal structure of Ce$_3$Pd$_{20}$Si$_6$ viewed along one of the cubic axes. The isosurfaces of magnetization density at the Ce2~(8c) position, predicted at the DFT/PBE/PAW level of theory, are shown in dark blue. (b)~The nearest-neighbor cluster ($T_d$ symmetry) around the magnetic Ce2 site, used to create a point-charge model in CASSCF modeling.\vspace{-4pt}}
\label{Fig:Theory}
\end{figure}

\vspace{-6pt}\subsection{Density functional theory (DFT) calculations}\vspace{-3pt}

Generally, no single-determinant model (DFT models included) can be considered reliable in describing 4\textit{f} and 5\textit{f} systems, which is explained by the multiconfigurational character of the full ground-state multiplet, even if the ground-state wave function can be formally represented by a single determinant. However, among all 4\textit{f} elements, the Ce$^{3+}$ ion is perhaps the only lanthanide where DFT models are capable of producing some single-particle properties reliably, in particular spin densities. In our previous studies of both metallic and insulating Ce-based systems like CeB$_6$~\cite{InosovAvdoshenko21} and KCeS$_2$~\cite{KulbakovAvdoshenko21}, we have shown that the accuracy of DFT in predicting the spin density distribution within the DFT/PBE/PAW framework compared to CASSCF-level calculations is surprisingly high.

\begin{table*}[t!]
\caption{Decomposition of SOC states of the low-energy multiplet of Ce$^{3+}$ on the 8c Wyckoff site in the $|J,m_J\rangle$ basis for pseudospin $\tilde{S}=1/2$ (doublets $D_1$--$D_3$) and for the combination of quartet ($\Gamma_8$) and the first excited doublet ($\Gamma_7$), including the corresponding $g$ tensors.}\label{Tab:CEF}
\begin{center}
\begin{tabular}{@{~}c@{~~}|@{~~}c@{~~}|@{~~}cccccc@{~~}|@{~~}lll@{~}}
\toprule
State & SOC energy & $|\text{\textpm 5/2}\rangle$ & $|\text{\textpm 3/2}\rangle$ & $|\text{\textpm 1/2}\rangle$ & $|\text{\txtmp 1/2}\rangle$ & $|\text{\txtmp 3/2}\rangle$ & $|\text{\txtmp 5/2}\rangle$ & \multicolumn{1}{c}{$g_x$}& \multicolumn{1}{c}{$g_y$} & \multicolumn{1}{c}{$g_z$}\\
                   & (meV) &  &  &  &  &  &  &  &  & \\
\midrule
$D_1$          & 0.0                       & --- & --- & 0.90 & 0.10 & --- & --- & 2.57 & 2.57 & 0.85 \\
$D_2$          & $4.0\cdot10^{-5}$ & 0.43 & 0.08 & --- & --- & 0.09 & 0.40 & 1.43 & 1.43 & 3.14 \\
$D_3$          & 3.7                       & --- & 0.83 & --- & --- & --- & 0.17 & 1.45 & 1.45 & 1.45 \\ \midrule
$\Gamma_8$ & 0.0                       & --- & --- & --- & --- & --- & --- & 1.00 & 1.00 & 1.00 \\
$\Gamma_7$ & 3.7                       & --- & 0.83 & --- & --- & --- & 0.17 & 1.45 & 1.45 & 1.45 \\
\bottomrule
\end{tabular}
\end{center}
\end{table*}

In the application to Ce$_3$Pd$_{20}$Si$_6$, we find, on the one hand, that the Ce ions at the 8c and 4a Wyckoff sites are characterized by almost identical Bader charges, with only about 5\% larger positive value at the 4a position. On the other hand, we find that the Bader basins have different sizes, with Ce at 4a having a 15\% smaller volume. This indicates that the 4\textit{f} electron at the 8c Wyckoff site is more localized compared to the 4a site, and the local physics of the 4\textit{f} shell typical for a Ce$^{3+}$ ion is engaged. We also note that further Bader analysis has revealed that local charge rearrangements make Si and Pd slightly positive and negative with $\Delta q = +0.6e$ and $-0.45e$, respectively.

\vspace{-6pt}\subsection{CASSCF model}\vspace{-3pt}

Using the resulting partial charges, we evaluated the CEF parameters and the local anisotropy of the magnetic Ce$^{3+}$ on the 8c Wyckoff site by quantum chemistry calculations at the CASSCF(1,7)/ANO-RCC-VDZ/RASSI-SO level of theory for a Ce$^{3+}$ ion placed in the $T_d$ point charge environment, as shown in Fig.~\ref{Fig:Theory}\,(b). An \textit{ab initio} calculation was done using \texttt{OpenMolcas} code~\cite{KresseHafner93, AquilanteAutschbach20} for the [CePd$_{16}$Si$_{12}$]$^{3+}$ model. We found that by using the effective charges for Pd and Si, derived from the Bader analysis, we underestimate the experimental CEF excitation observed in INS, which appears at 2.1~meV instead of the experimental value of 3.7~meV. Therefore, we scaled the value of both charges while keeping their ratio $|q(\text{Pd})/q(\text{Si})| = 1.34$ to bring the splitting between the $\Gamma_8$ ground state and the $\Gamma_7$ excited state to the experimental value. Formally analyzing the \textit{ab initio} spin-orbit ground state in terms of pseudospin $\tilde{S}=1/2$, we find the two lowest-energy doublets separated by only $4\cdot10^{-5}$~meV, with $D_1$ dominated by $|\text{\textpm}1/2\rangle$ and $D_2$ by $|\pmmp5/2\rangle$ projections of the total angular momentum $J$ of the $J=5/2$ multiplet. The resulting decomposition of single-ion SOC states in the $|J,m_J\rangle$ basis and the $g$-tensor structure for the lowest-energy atomic multiplet are summarized in Table~\ref{Tab:CEF}. Parameters of the SOC states and the derived \textit{ab initio} CEF parameters in Stevens-operator notation, $B_k^{\,q}$, are given in Table~\ref{Tab:Stevens} in the Appendix. In agreement with earlier studies~\cite{Pas08, Dee10}, the ground-state multiplet ($J=5/2$), represented by the quartet $\Gamma_8$ and doublet $\Gamma_7$, is dominated by high-order Stevens operators ($O_4$).

\begin{figure*}[bt]\vspace{-6pt}
\centerline{\includegraphics[width=\textwidth]{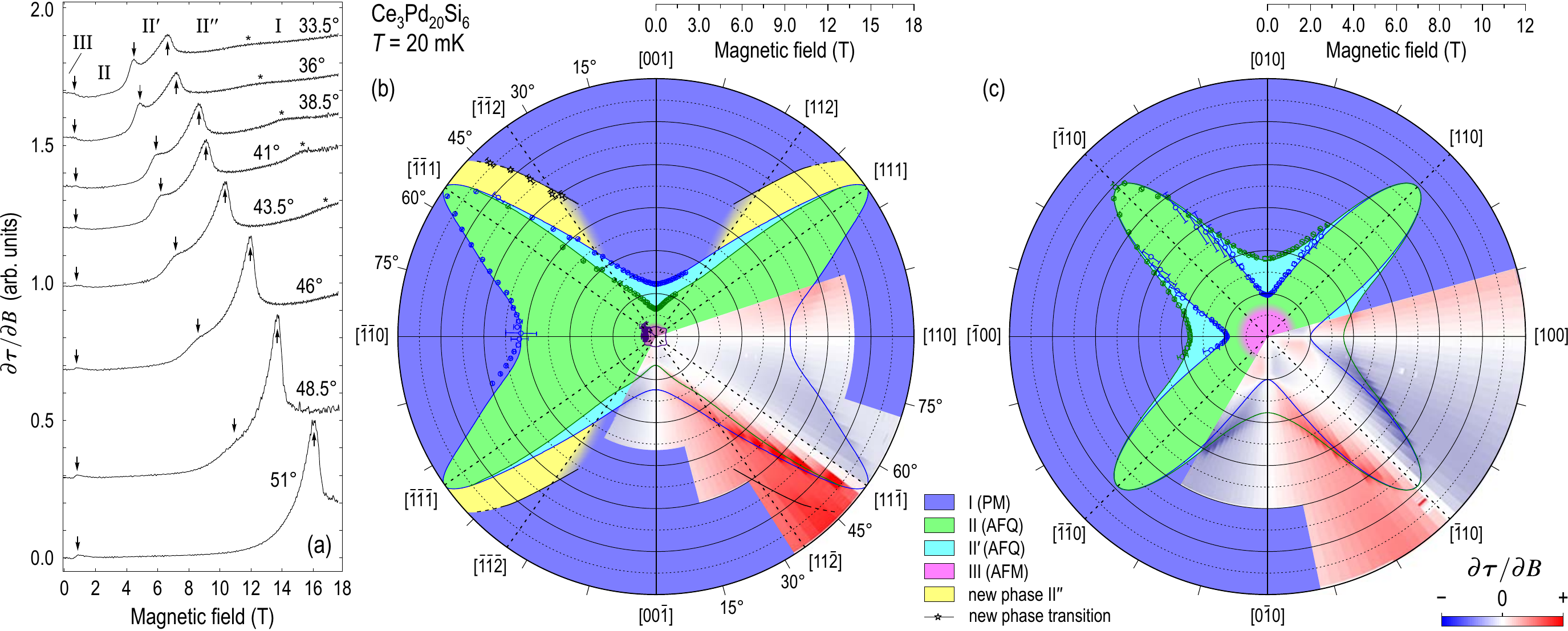}}\vspace{-2pt}
\caption{(a)~Magnetic-field derivative of the torque, $\partial\tau/\partial B$, measured at $T=20$~mK with torque magnetometry at different field angles in the $(1\overline{1}0)$ plane. The magnetic-field angles shown beside the curves are given with respect to the [001] cubic axis. Because of space reasons, only curves in the sector between the $[112]$ (35.3$^\circ$) and $[111]$ (54.7$^\circ$) directions are shown, shifted vertically for clarity. The arrows mark anomalies corresponding to the transitions between phases III, II, II$^\prime$, and II$^{\prime\prime}$. The weak anomaly at the upper boundary of phase~II$^{\prime\prime}$ is marked by asterisks. (b,\,c)~Field-angle magnetic phase diagrams of Ce$_3$Pd$_{20}$Si$_6$ at $T=20$~mK in polar coordinates, reconstructed from torque magnetometry data for fields rotated in the $(1\overline{1}0)$ and $(001)$ planes, respectively. The field derivative of the magnetic torque, $\partial\tau/\partial B$, composed of scans like those in panel (a), is plotted as a function of the magnetic field and its direction in the underlaid color map. Here red and blue colors correspond to the positive (+) and negative (--) signs of $\partial\tau/\partial B$, respectively. The stability regions for different phases are shaded, based on solid lines separating them that were fitted to the data points under constraints of the cubic crystal symmetry.\vspace{-2pt}}
\label{Fig:Torque}
\end{figure*}

\vspace{-6pt}\section{Field-angle phase diagram}\vspace{-1pt}\label{Sec:Magnetometry}\vspace{-5pt}

Torque magnetometry measurements in magnetic fields up to 18~T were carried out on a 5.0~mg single crystal of Ce$_3$Pd$_{20}$Si$_6$ at the National High Magnetic Field Laboratory using the setup described in Ref.~\citenum{InosovAvdoshenko21}. The sample, oriented with an x-ray Laue camera, was cut in two halves, the approximately equal pieces were glued to two cantilever paddles made of beryllium copper and mounted on a cylindrical sample holder placed inside the rotator with its rotation axis parallel to the $(1\overline{1}0)$ and $(001)$ crystal axes, respectively. The magnetic torque $\mathbold{\tau}=\mathbf{M}(\mathbf{B})\times\mathbf{B}$ that is exerted on the sample with magnetization $\mathbf{M}$ by an external magnetic field $\mathbf{B}$ was measured capacitively by the elastic deflection of the cantilever. The measurements were carried out in the 18/20~T superconducting magnet SCM1 equipped with a top-loading dilution refrigerator, at its base temperature of 20~mK. The rotation angle $\theta$ of the sample was controlled by a stepper motor with a resolution of 110 steps per 1$^\circ$ angle, giving the rotator approximately 0.02$^\circ$ of resolution. The samples were rotated in steps of 2.5$^\circ$, and field sweeps with a sweep rate between 0.2 and 0.3~T/min were carried out at every rotation step.

In Fig.\,\ref{Fig:Torque}\,(a), we show the magnetic-field derivative of the measured torque, $\partial\mathbold{\tau}(B,\theta)/\partial B$, which is equivalent to the angle derivative of the magnetization, $\partial M(B,\theta)/\partial\theta$~\cite{InosovAvdoshenko21}. The shown curves cover a range of angles between 33.5$^\circ$ and 51$^\circ$ in the $(1\overline{1}0)$ plane, as measured with respect to the [001] cubic axis. This approximately covers the sector between the [112] and [111] field directions, which is the range in which the weak anomaly corresponding to the high-field phase transition (marked by asterisks) is visible. To save space, the data in the rest of the measured angular range, as well as those for the $(001)$ field rotation plane, are only summarized in Figs.~\ref{Fig:Torque}\,(b) and \ref{Fig:Torque}\,(c) as color maps. Here the data are shown as polar plots with the radial and angular coordinates representing the magnitude and the direction of the magnetic field in the respective plane of rotation. The fields corresponding to phase transitions for every angle, estimated from the $\partial\mathbold{\tau}/\partial B$ curves as shown with arrows and asterisks in Fig.\,\ref{Fig:Torque}\,(a), are shown here with data~points.

Note that the AFM phase~III is essentially isotropic, as was known from earlier publications~\cite{MitamuraTayama10, OnoNakano13}. The transition from phase~III to phase~II can be seen only as a small step in $\partial\mathbold{\tau}/\partial B$ just below 1~T. In contrast, the transitions associated with the AFQ phases are much more pronounced and show strong dependence on the magnetic-field angle. The high-field boundary of phase II$^\prime$ can be seen as the dominant peak in $\partial\mathbold{\tau}/\partial B$, with the II-II$^\prime$ phase transition developing first as a shoulder and then as a standalone maximum on the low-field side of this peak as the field is rotated towards $[001]$. As one can see from Fig.\,\ref{Fig:Torque}, phase II$^\prime$ that was only known to exist for $\mathbf{B}\parallel[001]$ can actually be observed in a rather broad range of field angles up to $\theta \approx 50^\circ$ and 30$^\circ$ in the $(1\overline{1}0)$ and $(001)$ planes, respectively. In particular, it is very clearly pronounced for $\mathbf{B}\parallel[112]$ or equivalent directions. The two phase transitions merge as the field approaches either the [111] or the [110] axis, where phase II$^\prime$ vanishes.

Remarkably, at higher fields there is another anomaly in the $\partial\mathbold{\tau}/\partial B$ curves seen as a change in slope, which exhibits similarly strong dependence on the field direction. This transition is indicated with asterisks in Fig.\,\ref{Fig:Torque}\,(a) and with star symbols in Fig.\,\ref{Fig:Torque}\,(b). It can be seen in our measurements only in a narrow range of field angles in the $(1\overline{1}0)$ plane between the [112] and [111] field directions. While the corresponding anomaly in the magnetic torque is weaker than for the two other AFQ phase transitions, its clear dependence on the field angle suggests the existence of another high-field phase~II$^{\prime\prime}$ that escaped the attention in earlier studies, which concentrated only on the [001], [110], and [111] field directions, where this phase is not detectable or falls outside the accessible field range.

\vspace{-5pt}\section{Neutron scattering}\vspace{-3pt}\label{Sec:Neutrons}

To get an insight into the order-parameter symmetry and the excitation spectrum of all the mentioned phases, we performed neutron-scattering measurements at the cold-neutron triple-axis spectrometer \textsc{ThALES} (ILL, Grenoble, France)~\cite{RawILLdata21}. Two large single crystals of Ce$_3$Pd$_{20}$Si$_6$ with a combined mass of 5.9\,g were coaligned and mounted on a copper holder in the $\bigl(H\,K\,(H\!+\!K)/2\bigr)$ scattering plane, i.e. with the $[11\overline{2}]$ axis vertical. It was placed in a $^3$He/$^4$He dilution refrigerator inside a 15~T vertical-field cryomagnet, providing the lowest sample temperature of $\sim$50~mK. The spectrometer was operated with the fixed final neutron wavenumber $k_\text{f}=1.3$\,\AA$^{-1}$ and a cold beryllium filter installed between the sample and the analyzer to suppress higher-order contamination of the neutron beam. This experimental configuration is essentially the same as the one used in previously published measurements on the same sample for $\mathbf{B}\parallel[110]$ and $\mathbf{B}\parallel[001]$ field orientations~\cite{PortnichenkoPaschen16}. All data were collected at the base temperature.

\begin{figure}[b!]\vspace{-2pt}
\includegraphics[width=\columnwidth]{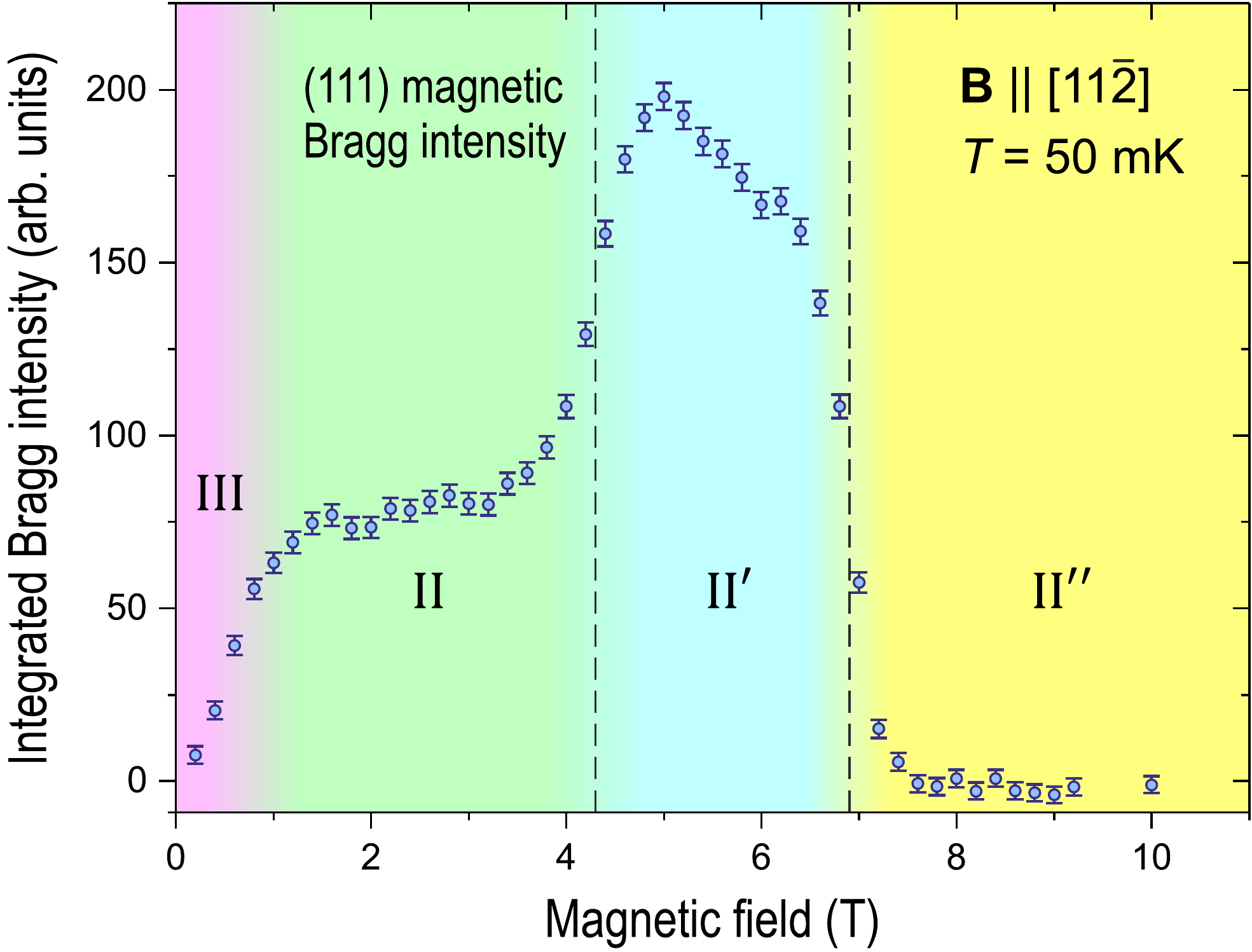}\vspace{-3pt}
\caption{Field-induced magnetic Bragg intensity $I_{\smash{(111)}}(B)$ at $\mathbf{Q}=(111)$, measured at $T=50$~mK as a function of the magnetic field applied along the $[11\overline{2}]$ direction (circles). The dashed lines show phase transitions between multipolar phases according to our magnetic torque data.\vspace{-4pt}}
\label{Fig:Bragg111vsB}
\end{figure}

\vspace{-3pt}\subsection{Elastic scattering}\vspace{-3pt}\label{SubSec:ElasticScattering}

\begin{figure*}[t!]\vspace{-8pt}
\centerline{\includegraphics[width=0.89\textwidth]{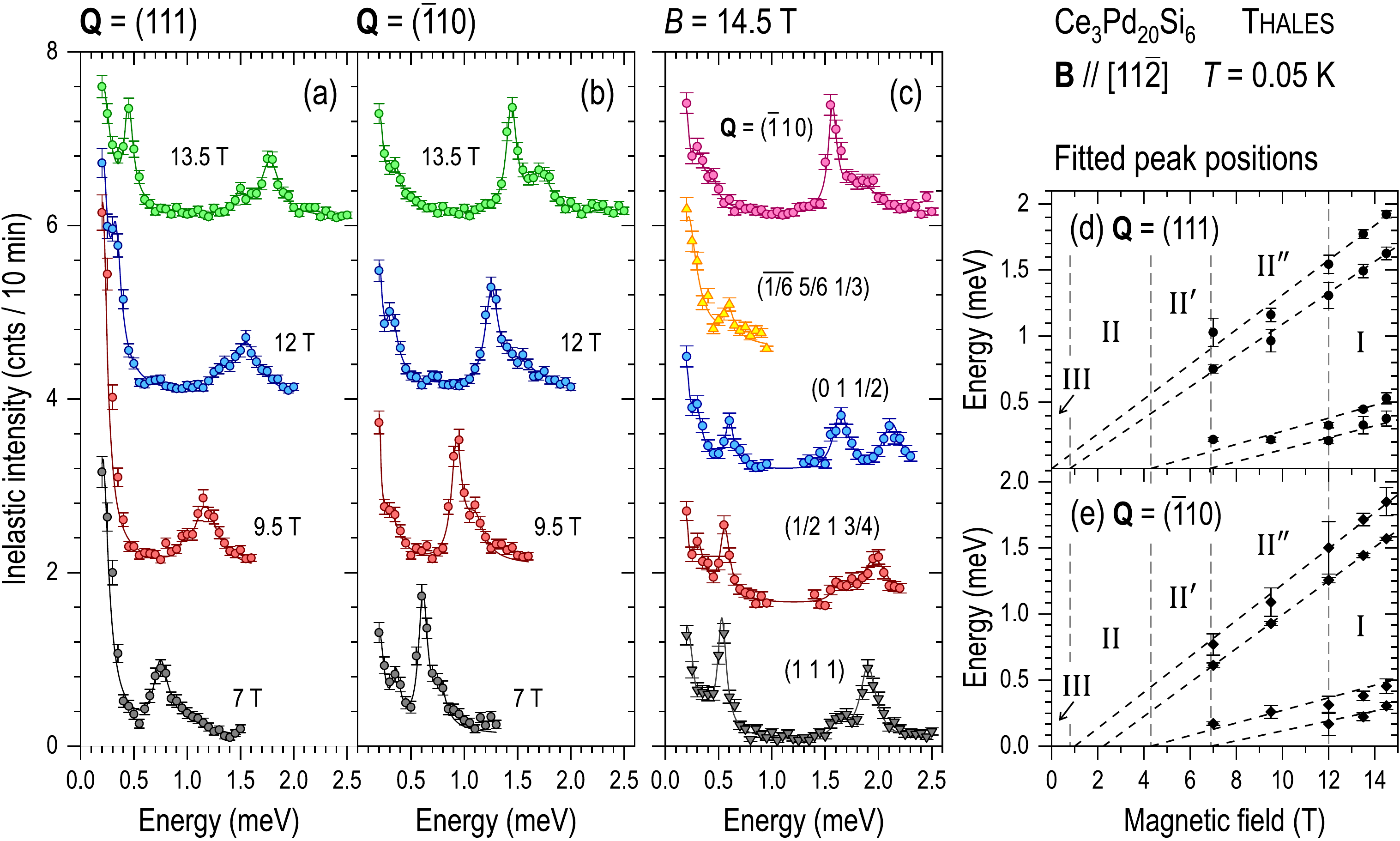}\vspace{-3pt}}
\caption{(a,\,b)~A selection of unprocessed INS data taken at $T = 0.05$~K at $\mathbf{Q}=(111)$ and $(\overline{1}10)$, respectively, in magnetic fields of 7, 9.5, 12, and 13.5~T applied along $[11\overline{2}]$. (c)~The data collected at the highest field of 14.5~T at different $\mathbf{Q}$ vectors from $(111)$ and $(\overline{1}10)$ as indicated beside each dataset. The data in panels (a\,--\,c) are offset vertically for clarity. (d,\,e)~Magnetic field dependence of the inelastic peak positions extracted from the fits shown in panels (a\,--\,c). The straight dashed lines are guides to the eyes, extrapolating the peak energies towards zero in smaller fields. Vertical dashed lines mark phase transitions for the $[11\overline{2}]$ field directions.\vspace{-1pt}}
\label{Fig:INS_RawData}
\end{figure*}

We start with presenting the results of elastic-scattering measurements. Here we benefit from the energy analysis to separate the weak elastic-scattering signal from inelastic contributions. To estimate the background-free Bragg intensity induced by the magnetic field $\mathbf{B}\parallel[11\overline{2}]$ at the $q_\text{II}=(111)$ wave vector, at every field value changed in 0.2~T steps we measured a momentum scan along $(1\!+\!h, 1\!-\!h, 1)$ at zero energy transfer. The nonmagnetic incoherent-scattering background and the field-independent Bragg scattering from the (111) structural reflection were eliminated by subtracting the fit of the zero-field measurement from all other datasets, as we already knew that the former contains no magnetic signal~\cite{PortnichenkoPaschen16}. The resulting peaks were well centered and showed no observable changes in shape within the whole measured field range apart from the sought intensity modulation. Therefore we could fit them globally with a Gaussian profile, sharing the background and peak width parameters for all scans, so that the peak amplitude remained as the only field-dependent parameter.

The resulting (111) magnetic peak intensity $I_{\smash{(\!1\!1\!1\!)}}(B)$ as a function of the magnetic field $\mathbf{B}\parallel[11\overline{2}]$ is shown in Fig.~\ref{Fig:Bragg111vsB} with circles. Already within phase~III, the peak amplitude immediately shows an approximately linear increase similar to that observed for $\mathbf{B}\parallel[110]$ and $\mathbf{B}\parallel[001]$ in our earlier work~\cite{PortnichenkoPaschen16}. Recalling that the (111) Bragg peak becomes visible because of dipolar moments induced by the magnetic field as a secondary order parameter on top of the ordered $O^0_2$ quadrupolar moments of the underlying AFQ phase, this indicates that the AFQ and AFM order parameters coexist. Note that the AFM magnetic Bragg peak at the ordering vector of phase~III, $q_\text{III}=(0\,0\,\frac{4}{5})$, persists to at least 0.6~T in a magnetic field independently of the field direction~\cite{PortnichenkoPaschen16}. In this field range and at the base temperature $T=50$~mK, the field-suppressed AFM peak at $q_\text{III}$ and the field-induced AFQ peak at $q_\text{II}$ are observed simultaneously. For comparison, the magnetic phase diagram of the closely related compound CeB$_6$ contains two AFM phases: a multi-$\mathbf{q}$ phase~III and a single-$\mathbf{q}$ phase III$^\prime$, separated by a field-driven phase transition. There, no field-induced magnetic scattering is seen at the AFQ wave vector within phase~III at base temperature~\cite{RossatMignod87, CameronFriemel16, JangPortnichenko17, InosovAvdoshenko21}, which proves that phase~III is of purely dipolar character. On the other hand, in CeB$_6$ the AFQ peak is induced by the field already in phase~III$^\prime$~\cite{JangPortnichenko17, InosovAvdoshenko21}, so both order parameters coexist just as in Ce$_3$Pd$_{20}$Si$_6$.

Upon suppression of phase~III, the intensity $I_{\smash{(\!1\!1\!1\!)}}(B)$ nearly flattens and remains practically constant in phase~II until approximately 3.5~T. Then, as the transition to phase II$^\prime$ is approached, the peak intensity rises again, reaching a maximum at 5~T. The center of this broadened step, that has a width of approximately 1~T, is centered at the transition from phase~II to phase II$^\prime$ extracted from torque magnetometry data for this field orientation, which is shown in Fig.~\ref{Fig:Bragg111vsB} with a dashed line at about 4.5~T. Note that the field-induced magnetic Bragg peak in phase~II$^{\prime}$, observed here with the maximal intensity, could not be seen previously for $\mathbf{B}\parallel[001]$, where $I_{\smash{(\!1\!1\!1\!)}}(B)$ dropped down to zero outside phase~II~\cite{PortnichenkoPaschen16}. This behavior also confirms the suggested $\Gamma_3^+$-type AFQ order for phase~II$^{\prime}$, implying that it remains magnetically hidden only for one single magnetic field direction. For any arbitrary field angle deviating from the $\langle001\rangle$ cubic axes, dipolar moments visible to neutron scattering are induced, as expected from the theory~\cite{ShiinaShiba97, ShiinaSakai98} that predicts a maximum in the total field-induced dipolar moment for the $\langle112\rangle$ magnetic-field directions.

At higher fields, on crossing the second multipolar transition from phase~II$^\prime$ to phase~II$^{\prime\prime}$, the (111) magnetic peak intensity is suppressed to zero. The width of this transition is also on the order of 1~T, and it is centered at approximately 7~T in agreement with the magnetic torque measurements. In the absence of Bragg scattering in phase~II$^{\prime\prime}$, we cannot determine its propagation vector and order parameter symmetry from elastic scattering in the present configuration. This phase could represent another type of long-range order of some kind of a spin-nematic state. To reveal the collective excitations that could shed light on the possible ordering vectors in this phase, we now turn to the discussion of our INS results.

\vspace{-5pt}\subsection{Inelastic scattering}\vspace{-5pt}

The INS measurements were carried out in the same configuration of the \textsc{ThALES} spectrometer with $k_\text{f}=1.3$\,\AA$^{-1}$ and the crystal mounted in the $\bigl(H\,K\,(H\!+\!K)/2\bigr)$ scattering plane. Typical spectra taken at $\mathbf{Q}=(111)$ and $(\overline{1}10)$ wave vectors at $T=0.05$~K in different magnetic fields between 7 and 13.5~T, as well as the spectra at different wave vectors between $(111)$ and $(\overline{1}10)$, taken at the highest magnetic field of 14.5~T, are shown in Figs.~\ref{Fig:INS_RawData}\,(a), \ref{Fig:INS_RawData}\,(b), and \ref{Fig:INS_RawData}\,(c), respectively. The spectra are fitted by a sum of constant nonmagnetic background, incoherent elastic line in the form of a Gaussian, and up to 4 inelastic peaks described by the Lorentzian line shape~\cite{RegnaultErkelens88, Hewson93, Robinson00},\vspace{-3pt}
\begin{multline}\label{Eq:Quasielastic}
S(\mathbf{Q},\omega)\propto F^2(\mathbf{Q})\,\frac{\chi_0(\mathbf{Q})}{1-\exp(-\hslash\omega/k_\text{B}T)}\\
\times\frac{\omega}{2\pi}\biggl(\frac{\Gamma}{\hslash^2(\omega-\omega_0)^2+\Gamma^2}+\frac{\Gamma}{\hslash^2(\omega+\omega_0)^2+\Gamma^2}\biggr),
\end{multline}
where $F(\mathbf{Q})$ is the Ce$^{3+}$ magnetic form factor, $\chi_0(\mathbf{Q})$ is the momentum-dependent static susceptibility, and $\Gamma$ is the half-width of the Lorentzian centered at $\pm\hslash\omega_0$.

\begin{figure}[t!]\vspace{3pt}
\includegraphics[width=\columnwidth]{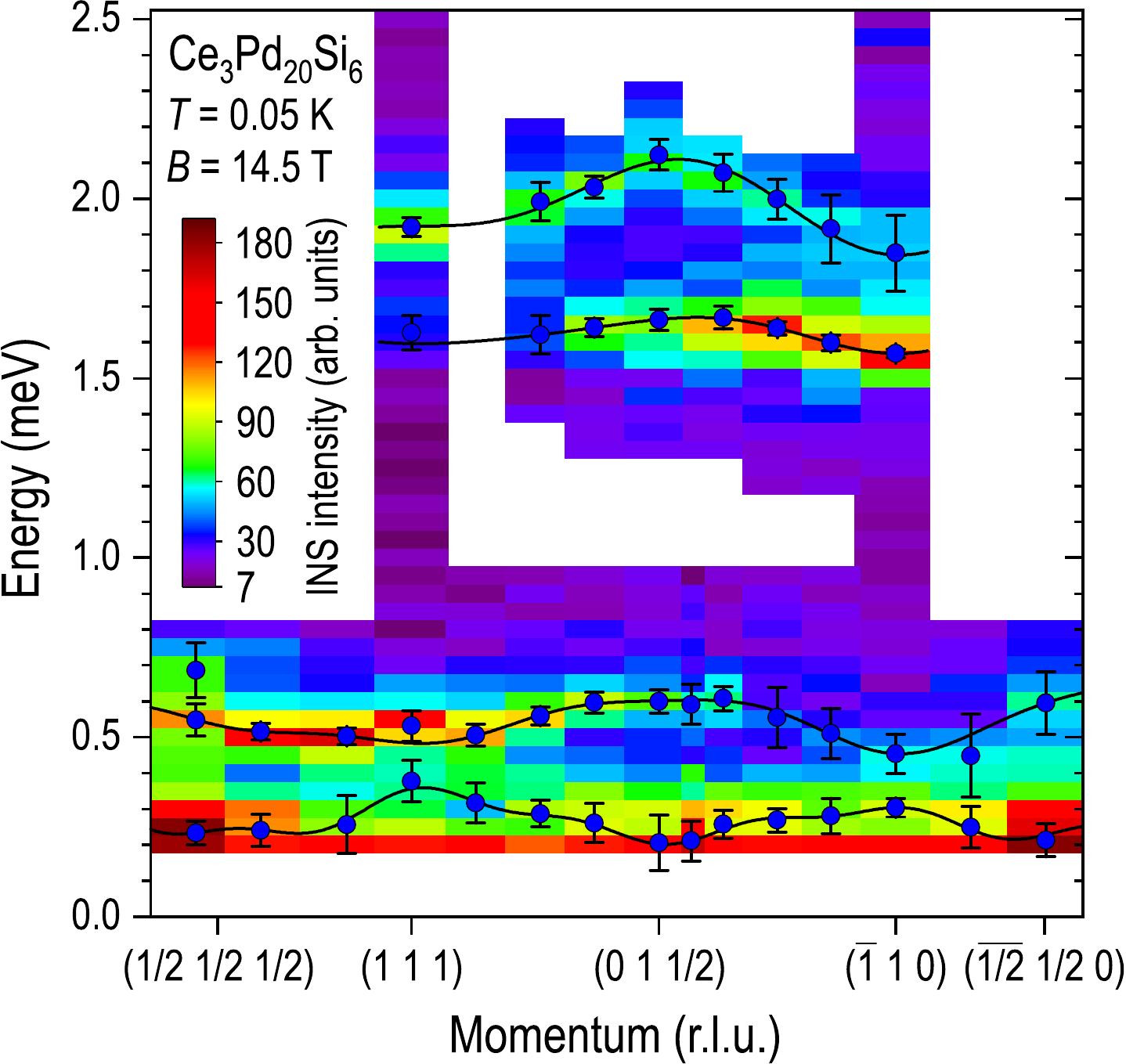}
\caption{Dispersion of field-induced magnetic excitations in Ce$_3$Pd$_{20}$Si$_6$ measured in a magnetic field of 14.5~T applied along the $[11\overline{2}]$ axis. The color map is composed of energy scans such as those in Fig.~\ref{Fig:INS_RawData}\,(c), measured along high-symmetry directions following the polygonal path $(0.4~0.4~0.4)$--$(111)$--$(\overline{1}\,1\,0)$--$(\overline{0.5}~0.5~0)$. The fitted peak position are shown as data points, the lines are an empirical fit to the data.}
\label{Fig:colormap}
\end{figure}

Already in the raw data, one can see a gradual shift in the peak positions toward higher energies with increasing magnetic field. We also observe the appearance of an increasing number of new peaks induced at high fields. Indeed, as we know from earlier measurements, the spectrum in zero field is described by a single broad Lorentzian centered at an energy smaller than the peak width~\cite{PortnichenkoCameron15}. If the magnetic field is applied along one of the $\langle110\rangle$ directions for which the phase diagram shows only a single multipolar phase~II, just a single inelastic line is induced by the field~\cite{PortnichenkoNikitin19}. In contrast, if the field is applied along one of the $\langle001\rangle$ directions, where both phase~II and phase~II$^\prime$ are present, at least two dispersive excitations can be clearly observed (modes $\hbar\omega_1$ and $\hbar\omega_2$ in Ref.~\citenum{PortnichenkoNikitin19}). Our present data for $\mathbf{B}\parallel[11\overline{2}]$ reveal four different field-induced modes in high magnetic fields. Note that the two lowest-energy excitations are not present in phase~II but only appear in phase~II$^\prime$ or phase~II$^{\prime\prime}$. The dispersion of all four field-induced excitations across the first Brillouin zone can be clearly seen in the color map of INS intensity in Fig.~\ref{Fig:colormap}. We therefore conclude that there is a close relationship between the number of nondegenerate field-induced collective excitations and that of multipolar phases in the magnetic phase diagram of Ce$_3$Pd$_{20}$Si$_{6}$.

To illustrate how new peaks appear upon increasing magnetic field, in Figs.~\ref{Fig:INS_RawData}\,(d) and \ref{Fig:INS_RawData}\,(e) we show the field dependence of the fitted peak positions at $\mathbf{Q}=(111)$ and $(\overline{1}10)$, respectively. By extrapolation to zero energy (dashed lines), one can estimate the magnetic fields at which these excitations emerge. Apparently, only the two highest-energy modes are observed in phase~II, whereas the two low-energy modes appear at the onsets of phase II$^\prime$ and phase II$^{\prime\prime}$, respectively. These lower-energy excitations are characterized by a considerably smaller slope ($g$ factor) in the magnetic field dependence. Our observation suggests that every field-driven phase transition from one multipolar phase to another brings about a new excitation mode that starts from zero at that phase transition, hence we can juxtapose each phase with its own low-energy excitation that gets softened and disappears as this phase is suppressed with decreasing magnetic field.

It is natural to expect that the dispersion of the lowest-lying excitation in a given phase has a minimum at the respective ordering vector, representing the Goldstone mode of the corresponding order parameter. Indeed, as one can see in Fig.~\ref{Fig:colormap}, all three higher-energy excitations that emerge in phases II and II$^\prime$ exhibit local dispersion minima at the $(111)$ point, where field-induced Bragg intensity was observed in Sec.~\ref{SubSec:ElasticScattering}. In contrast, the lowest-energy excitation emerging in phase~II$^{\prime\prime}$ exhibits a local maximum at $\mathbf{Q}=(111)$, while the minima appear at the $(\frac{1}{2}\frac{1}{2}\frac{1}{2})$, $(\frac{\overline{1}}{2}\frac{1}{2}0)$ and $(\frac{\overline{1}}{6}\frac{5}{6}\frac{1}{3})$ wave vectors. Note that the latter represents the orthogonal projection of the $(010)$ vector onto our scattering plane. These three wave vectors are therefore plausible candidates for the ordering vector of phase~II$^{\prime\prime}$, as within our experimental resolution the energies of the three minima cannot be conclusively distinguished.

Because the $(010)$ wave vector lies out of the scattering plane, it is impossible to verify the existence of a field-induced Bragg peak at this position in phase II$^{\prime\prime}$ in the present configuration. This would be also impossible with a horizontal-field magnet at ILL, where magnetic field is limited to 4~T. The only option is to measure this wave vector with time-of-flight (TOF) neutron spectroscopy in a vertical magnetic field of 10~T (which is to our knowledge the highest magnetic field currently available at cold-neutron TOF instruments). While this challenging experiment is beyond the scope of our present work, it should be considered in the future.

\vspace{-5pt}\section{Summary and Discussion}\vspace{-5pt}

\begin{figure*}[bt]\vspace{-6pt}
\centerline{\includegraphics[width=\textwidth]{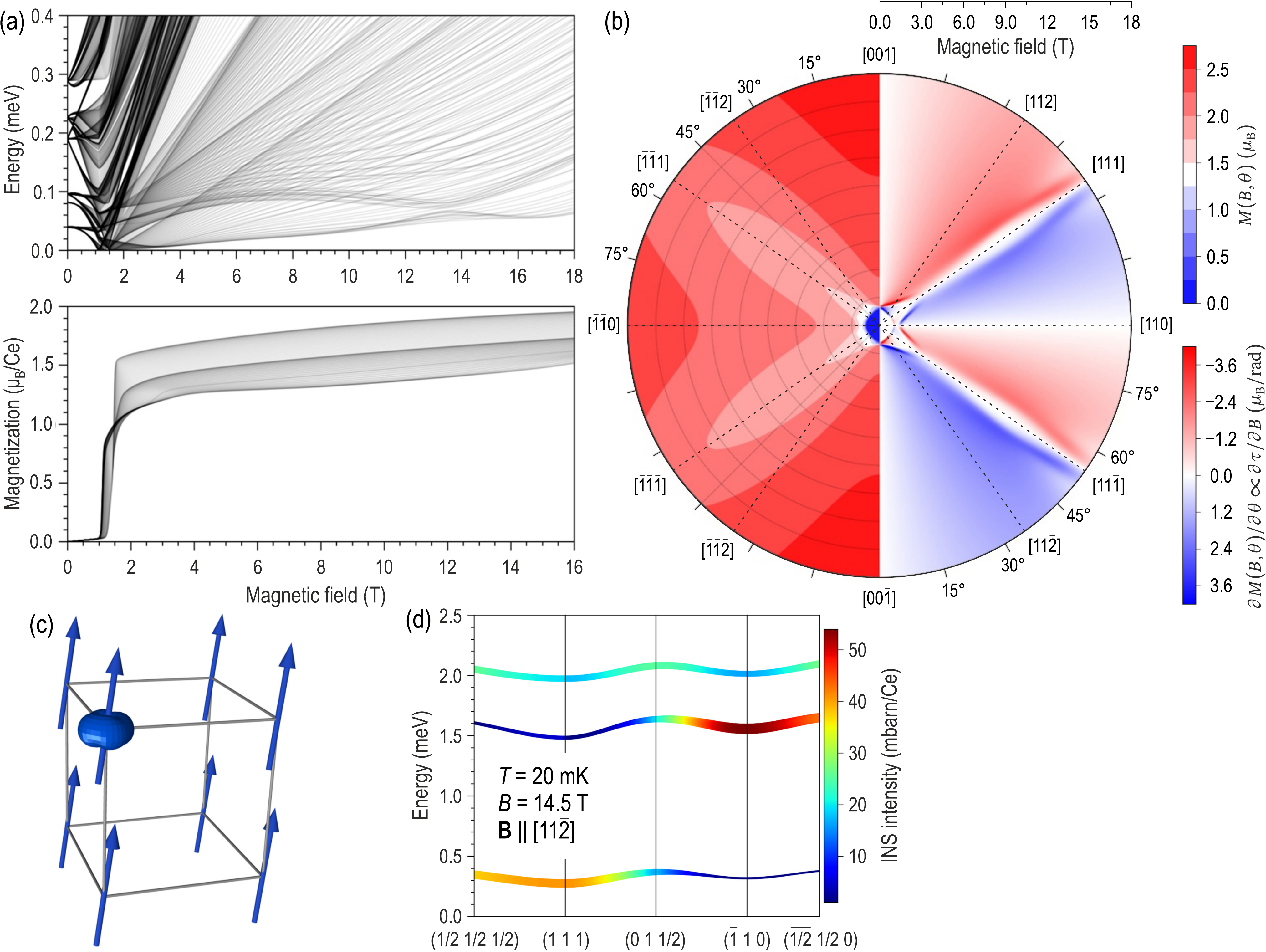}}
\caption{(a)~Superimposed Zeeman diagrams (top) and magnetization curves (bottom), calculated in the two-ion model as a function of magnetic field $\mathbf{B}$ for different field directions uniformly distributed in the $(110)$ plane. (b)~Contour map of the calculated magnetization $M(B, \theta)$ (left) and a color map of its angular derivative $\partial M/\partial\theta$ (equivalent to the experimentally measured $\partial\tau/\partial B$), presented in polar coordinates vs. magnetic-field strength up to 18~T and rotation angle $\theta$ in the $(1\overline{1}0)$ plane. (c)~Field-polarized magnetic state at $T=50$~mK and $B=14.5$~T for $\mathbf{B}\parallel[11\overline{2}]$. (d)~Dispersion of the field-induced magnetic excitations at $B=14.5$~T, plotted along the high-symmetry polygonal path in $\mathbf{Q}$ space for $\mathbf{B}\parallel[11\overline{2}]\perp\mathbf{Q}$. The color shading represents the expected INS intensities, to be compared with the experimental data in Fig.~\ref{Fig:colormap}.}
\label{Fig:Theory2}
\end{figure*}

In summary, we used a combination of magnetic torque measurements and neutron scattering to reveal a high-field phase transition in Ce$_3$Pd$_{20}$Si$_6$ that appears only for certain field directions in the vicinity of the $\langle112\rangle$ directions but not for fields along the $\langle001\rangle$ or $\langle110\rangle$ directions, which explains why this phase was overlooked in earlier works. In total, this results in three distinct hidden-order phases in this compound, separated by two field-driven quantum phase transitions. Contrary to the AFM phase whose stability region is almost isotropic with respect to the field direction, the multipolar phases show strong field-direction anisotropy revealed in the torque magnetometry. Not only the critical fields, but also the number of field-induced phases depends on the magnetic field direction.

Using elastic neutron scattering in a magnetic field applied along the $[11\overline{2}]$ axis, we could observe field-induced magnetic Bragg scattering at the $(111)$ wave vector in phases II and II$^\prime$. Comparing this result with the previously reported absence of Bragg intensity in phase II$^\prime$ for $\mathbf{B}\parallel[001]$~\cite{PortnichenkoPaschen16}, we confirm the suggestion that phases II and II$^\prime$ represent AFQ order characterized by the same wave vector but different ordered quadrupolar moments: $\Gamma_3^+$-type in phase~II and $\Gamma_5^+$-type in phase~II$^\prime$, separated by a multipolar phase transition. The second phase transition from phase II$^\prime$ to phase II$^{\prime\prime}$ is suggested by the magnetic torque measurements, yet no field-induced magnetic Bragg intensity could so far be observed in phase~II$^{\prime\prime}$. While this does not exclude that such intensity may appear at some wave vector not covered in our measurements, dispersion minima of the lowest-lying magnetic excitation associated with phase~II$^{\prime\prime}$ suggest three possible candidates for the ordering vector in this phase: $(\frac{1}{2}\frac{1}{2}\frac{1}{2})$, $(\smash{\frac{\overline{1}}{2}}\frac{1}{2}0)$, and $(010)$. An alternative scenario is that phase~II$^{\prime\prime}$ may represent a multipolar analog of a spin-nematic phase that does not break translational symmetry, by analogy with putative spin-nematic phases observed at high fields in frustrated magnets~\cite{SvistovFujita11, FeleaYasin12, KohamaIshikawa19, SkoulatosRucker19}, which also show no signatures in magnetic Bragg scattering but are evidenced by an undersaturated magnetization or anomalies in ultrasound velocity or in specific heat.

While the microscopic nature of the order parameter in phase~II$^{\prime\prime}$ remains to be clarified, here we observed field-induced magnetic excitations associated with phases II, II$^\prime$, and II$^{\prime\prime}$. The energy of magnetic excitations generally increases with magnetic field across the Brillouin zone, and a new dispersive band of collective excitations appears at every phase transition, which results in a total of 4 distinct collective modes for $\mathbf{B}\parallel[11\overline{2}]$ within phase II$^{\prime\prime}$. As we observed only one such excitation for $\mathbf{B}\parallel[001]$ and two excitations for $\mathbf{B}\parallel[1\overline{1}0]$ in an earlier work~\cite{PortnichenkoNikitin19}, we must conclude that some excitation branches become degenerate for high-symmetry field directions. This emphasizes the importance of INS measurements in arbitrary field directions for understanding all relevant degrees of freedom in the system. Our results also suggest a direct correspondence between the number of distinct phases in the magnetic phase diagram for a particular field direction and that of nondegenerate collective modes induced by the magnetic field.

The lowest-lying excitation associated with phase~II$^{\prime\prime}$ has a qualitatively different dispersion with different positions of local minima in $\mathbf{Q}$ space, compared to that of the three other excitation branches that exist already in phases II and II$^\prime$. This is a strong indication for a change in the magnetic ordering vector across the transition from phase~II$^\prime$ to phase~II$^{\prime\prime}$. As we noted earlier~\cite{PortnichenkoNikitin19}, in the absence of Bragg scattering, the dispersion of field-induced magnetic excitations remains the only measurable quantity from which indirect conclusions about the symmetry of the underlying multipolar order parameter of a hidden-order phase can be reached. Our present observation of multiple excitation branches in Ce$_3$Pd$_{20}$Si$_6$ is an excellent illustration of this principle. As soon as a reliable theoretical description of the field-induced multipolar phases and their excitations becomes available, our data obtained for different field directions will serve as a stringent test case for validating the theory or fitting its free parameters to the experiment.

Some initial qualitative understanding of the field-direction anisotropy in Ce$_3$Pd$_{20}$Si$_6$ and the behavior of field-induced magnetic excitations can be obtained with the minimal phenomenological model developed previously for CeB$_6$~\cite{InosovAvdoshenko21}. It involves two nearest-neighbor Ce$^{3+}$ sites in identical CEF environments, coupled by just one nearest-neighbor isotropic exchange interaction $\mathcal{j}_{12}$. The Hamiltonian of such a two-site model takes the form
\begin{equation}\label{Eq:Hamiltonian}
\hamcal_\text{m2}=\hamcal_{\text{CF}_\text{Ce1}} + \hamcal_{\text{CF}_\text{Ce2}} + \hamcal_\text{Zee} -2\mathcal{j}_{12\,}^{\phantom{~}}{\hat{J\kern2.3pt}\kern-3pt}_{\text{Ce1}}{\hat{J\kern2.3pt}\kern-3pt}_{\text{Ce2}},
\end{equation}
where $\hamcal_{\text{CF}_\text{Ce1,2}}$ are the single-ion CF Hamiltonians for the two Ce sites, described by the corresponding \textit{ab initio} crystal-field parameters $B_k^{\,q}$ in Stevens notation,
\begin{equation}\label{Eq:Stevens}
\hamcal_{\text{CF}_\text{Ce1,2}} =\kern-3pt\mathlarger{\sum}_{k=2,4,6}\,\sum_{q=-k}^k{\kern-3pt}B^{\,q}_k\hat{O}^{\,q}_k,
\end{equation}
and $\hamcal_\text{Zee}=\mu_\text{B}\hat{g}\cdot\hat{J\kern2.3pt}\kern-3pt\cdot\mathbf{B}$ is the Zeeman energy. The last term in Eq.~\eqref{Eq:Hamiltonian} stands for the exchange energy between the nearest-neighbor localized moments ${\hat{J\kern2.3pt}\kern-3pt}_{\rm Ce1}$ and ${\hat{J\kern2.3pt}\kern-3pt}_{\rm Ce2}$. The isotropic exchange of $\mathcal{j}_{12}=-0.015$~meV satisfies the overall anisotropy of the field-angle magnetic phase diagrams and some magnetic excitations. For example, Fig.~\ref{Fig:Theory2}\,(a) shows magnetic-field dependence of CEF state energies, measured relative to the lowest-energy state at the corresponding $\mathbf{B}$ value, superimposed for different field directions with a uniform distribution in $\theta$ within the $(110)$ plane. The combination of ground-state quartets on the two Ce$^{3+}$ ions results in 16 eigenstates whose degeneracy is partly removed by the intersite interaction even in zero magnetic field. Upon the application of a magnetic field, the remaining degeneracy is lifted due to the Zeeman splitting, resulting in a crowded Zeeman diagram with a strong dependence on the magnetic field direction that results from the anisotropy of $g$ tensors of the respective SOC states. The Zeeman diagram in Fig.~\ref{Fig:Theory2}\,(a) shows a level crossing at $B=1.0$--1.2~T (depending on the field direction), which corresponds to a step in the on-site magnetization in Fig.~\ref{Fig:Theory2}\,(b) and coincides with the transition point between phases III and II. There is also a level crossing around $B=13.0$--13.5~T in the lowest-energy excited state, in the approximate field range where the high-field boundary of the AFQ phases is experimentally observed. The calculated magnetic field dependence of the on-site magnetization and its derivative with respect to the field angle, $\partial M(B,\theta)/\partial\theta$, which is equivalent to the magnetic-field derivative of the torque, are shown in Fig.~\ref{Fig:Theory2}\,(b). There is also a qualitative similarity between the anisotropy in $\partial M(B,\theta)/\partial\theta$ in our two-site model and the experimental phase diagram in Fig.~\ref{Fig:Torque}\,(b), both showing elongated lobes extending along the $\langle111\rangle$ directions.

Using the nearest-neighbor interactions in a periodic setting for the magnetic subunit and self-consistent calculation for the magnetic moment configurations search as implemented in \textsc{McPhase}~\cite{Rotter04}, we optimized the magnetic ordered state at $T=50$~mK and $B=14.5$~T for $\mathbf{B}\parallel[11\overline{2}]$. The orientation of magnetic moments in the unit cell and spin density are shown in Fig.~\ref{Fig:Theory2}\,(c). The field-induced magnetic excitations shown in Fig.~\ref{Fig:Theory2}\,(d) were computed using the \textit{mcdisp} module in \textsc{McPhase}. They consist of three branches that exhibit remarkable similarity to the three high-energy modes observed experimentally (see Fig.~\ref{Fig:colormap}). The maxima and minima of the dispersion, as well as the dynamical structure factor, are qualitatively reproduced by the calculation. However, the lowest-energy excitation in the INS data is not captured by the model.

The level of agreement between experiment and our oversimplified two-site model is actually surprising. The experimental features observed in torque magnetometry correspond to phase transitions between long-range ordered phases, while the theoretical calculations are based on a local model that is in principle unable to describe long-range order. We therefore conclude that the field-direction anisotropy of the ordered phases must have local origin, mimicking the anisotropy of transitions between local CEF ground states of the interacting Ce$^{3+}$ sites.\vspace{-7pt}

\begin{acknowledgments}
We thank P.~Thalmeier for stimulating discussions and helpful suggestions and acknowledge technical support from E.~Villard and the ILL sample environment team during the \textsc{ThALES} experiment. This study was funded in part by the German Research Foundation (DFG) under the individual research grants \mbox{PO~2621/1-1}, \mbox{AV~169/3-1}, and by the W\"urzburg-Dresden Cluster of Excellence on Complexity and Topology in Quantum Matter\,---\,\textit{ct.qmat} (EXC 2147, project-id 390858490). The authors from Vienna acknowledge financial support from the Austrian Science Fund (FWF, projects P29296-N27 and I5868-N\,--\,FOR 5249, QUAST). A portion of this work was performed at the National High Magnetic Field Laboratory, which is supported by the National Science Foundation Cooperative Agreement No.~DMR-1644779 and the state of Florida.
\end{acknowledgments}

\vspace{-5pt}\section*{Appendix}\vspace{-5pt}

\subsection*{Theoretical methods and models}

\paragraph*{DFT model.} The structure was optimized and the ground-state wave function analyzed at the DFT/PBE/PAW level of theory using projector augmented-wave method as implemented in the \texttt{VASP}~v.5 code and the standard pseudopotential \cite{KresseHafner93, ZirngieblHillebrands84, ZirngieblHillebrands85}.

\begin{table}[b!]\vspace{-3pt}
\caption{Parameters of the SOC states for the [CePd$_{16}$Si$_{12}$]$^{3+}$ cluster at the DKH2/CAS(1,7)/RASSI-SO/VDZ-RCC level of theory.\vspace{-1pt}}\label{Tab:SOC}
\begin{center}
\begin{tabular}{l@{~~~}c@{\qquad}l@{~~~}c}
\toprule
SOC-id & Energy (eV) & Energy (cm$^{-1}$) & $J$-block \\
\midrule
1, 2 & {\tt 0.0000000000} & {\tt ~~~0.0000} & {\tt 2.5} \\
3, 4 & {\tt 0.0000000398} & {\tt ~~~0.0003} & {\tt 2.5} \\
5, 6 & {\tt 0.0036727506} & {\tt ~~29.6227} & {\tt 2.5} \\
7, 8 & {\tt 0.3009323650} & {\tt 2427.1832} & {\tt 3.5} \\
9, 10 & {\tt 0.3009324218} & {\tt 2427.1837} & {\tt 3.5} \\
11, 12 & {\tt 0.3010185962} & {\tt 2427.8787} & {\tt 3.5} \\
13, 14 & {\tt 0.3062391716} & {\tt 2469.9855} & {\tt 3.5} \\
\bottomrule
\end{tabular}\vspace*{-8pt}
\end{center}
\end{table}

\paragraph*{\textit{Ab initio} method.} The first-principles CASSCF calculations were performed at the DKH2/CAS(1,7)/RASSI-SO/VDZ-RCC level using \texttt{OpenMolcas}~\cite{RoosLindh08, AquilanteAutschbach20} software. The complete active space (CAS) method was employed to treat a single electron on seven 4\textit{f} orbitals with total spin $S = 1/2$. The spin-free CAS solutions were used in further state interaction modeling using the SOC Hamiltonian (Table~\ref{Tab:SOC}). Furthermore, for all doubly degenerate SOC states, the $g$ tensors were computed using first-order perturbation theory. The SOC states were projected on the $J = 5/2$ multiplet using \textit{ab initio} CEF, derived using the \texttt{SINGLE\_ANISO} module~\cite{ChibotaruUngur12}.

\begin{table}[h]\vspace{-3pt}
\caption {$B_k^{\,q}$ parameters in Stevens-operator notation for the crystal-field Hamiltonian of the $J=\text{5/2}$ multiplet.\vspace{-4pt}} \label{Tab:Stevens}
\begin{center}
\begin{tabular}{l@{\quad}r@{\quad}r}
\toprule
$k$ & $q$ & \multicolumn{1}{c}{$B_k^{\,q}$ (cm$^{-1}$)} \\
\midrule
{\tt 2} & {\tt -2} & {\tt 0.15253489841400E-07} \\
{\tt 2} & {\tt -1} & {\tt -0.18705241307101E-04} \\
{\tt 2} & {\tt  0} & {\tt  -0.24429880473422E-02} \\
{\tt 2} & {\tt  1} & {\tt 0.72596761009504E-04} \\
{\tt 2} & {\tt  2} & {\tt -0.87364137479673E-08} \\
\midrule
{\tt 4} & {\tt -4} & {\tt  0.13556694960268E+00} \\
{\tt 4} & {\tt -3} & {\tt  0.56906727585488E-04} \\
{\tt 4} & {\tt -2} & {\tt  0.91081049908721E-07} \\
{\tt 4} & {\tt -1} & {\tt -0.17027203720282E-03} \\
{\tt 4} & {\tt  0} & {\tt -0.82449053000277E-01} \\
{\tt 4} & {\tt  1} & {\tt  0.66056580524203E-03} \\
{\tt 4} & {\tt  2} & {\tt -0.25628763149983E-06} \\
{\tt 4} & {\tt  3} & {\tt -0.67196521793959E-03} \\
{\tt 4} & {\tt  4} & {\tt -0.38722641278742E+00} \\
\bottomrule\vspace{-14pt}
\end{tabular}
\end{center}
\end{table}

\bibliographystyle{my-apsrev}\bibliography{Ce3Pd20Si6}\vspace*{-3pt}

\end{document}